\documentclass[11pt,twoside]{article}
\usepackage{asp2014}

\aspSuppressVolSlug
\resetcounters

\begin{document}

\title{Fundamentals of effective cloud management for the new NASA Astrophysics Data System}

\author{Sergi~Blanco-Cuaresma,
        Alberto~Accomazzi,
        Michael~J.~Kurtz,
        Edwin~Henneken,
        Carolyn~S.~Grant,
        Donna~M.~Thompson,
        Roman~Chyla,
        Stephen~McDonald,
        Golnaz~Shapurian,
        Timothy~W.~Hostetler,
        Matthew~R.~Templeton,
        Kelly~E.~Lockhart,
        Kris~Bukovi,
        and Nathan~Rapport
  }
\affil{Harvard-Smithsonian Center for Astrophysics, 60 Garden Street, Cambridge, MA 02138, USA \email{sblancocuaresma@cfa.harvard.edu}}

\paperauthor{Sergi~Blanco-Cuaresma}{sblancocuaresma@cfa.harvard.edu}{0000-0002-1584-0171}{Harvard-Smithsonian Center for Astrophysics}{HEAD}{Cambridge}{MA}{02138}{USA}
\paperauthor{Alberto~Accomazzi}{aaccomazzi@cfa.harvard.edu}{0000-0002-4110-3511}{Harvard-Smithsonian Center for Astrophysics}{HEAD}{Cambridge}{MA}{02138}{USA}
\paperauthor{Michael~J.~Kurtz}{kurtz@cfa.harvard.edu}{0000-0002-6949-0090}{Harvard-Smithsonian Center for Astrophysics}{HEAD}{Cambridge}{MA}{02138}{USA}
\paperauthor{Edwin~A.~Henneken}{ehenneken@cfa.harvard.edu}{0000-0003-4264-2450}{Harvard-Smithsonian Center for Astrophysics}{HEAD}{Cambridge}{MA}{02138}{USA}
\paperauthor{Carolyn~S.~Grant}{cgrant@cfa.harvard.edu}{0000-0003-4424-7366}{Harvard-Smithsonian Center for Astrophysics}{HEAD}{Cambridge}{MA}{02138}{USA}
\paperauthor{Donna~M.~Thompson}{dthompson@cfa.harvard.edu}{0000-0001-6870-2365}{Harvard-Smithsonian Center for Astrophysics}{HEAD}{Cambridge}{MA}{02138}{USA}
\paperauthor{Roman~Chyla}{rchyla@cfa.harvard.edu}{0000-0003-3041-2092}{Harvard-Smithsonian Center for Astrophysics}{HEAD}{Cambridge}{MA}{02138}{USA}
\paperauthor{Stephen~McDonald}{stephen.mcdonald@cfa.harvard.edu}{0000-0003-1270-0605}{Harvard-Smithsonian Center for Astrophysics}{HEAD}{Cambridge}{MA}{02138}{USA}
\paperauthor{Golnaz~Shapurian}{gshapurian@cfa.harvard.edu}{0000-0001-9759-9811}{Harvard-Smithsonian Center for Astrophysics}{HEAD}{Cambridge}{MA}{02138}{USA}
\paperauthor{Timothy~W.~Hostetler}{thostetler@cfa.harvard.edu}{0000-0001-9238-3667}{Harvard-Smithsonian Center for Astrophysics}{HEAD}{Cambridge}{MA}{02138}{USA}
\paperauthor{Matthew~R.~Templeton}{matthew.templeton@cfa.harvard.edu}{0000-0003-1918-0622}{Harvard-Smithsonian Center for Astrophysics}{HEAD}{Cambridge}{MA}{02138}{USA}
\paperauthor{Kelly~E.~Lockhart}{kelly.lockhart@cfa.harvard.edu}{0000-0002-8130-1440}{Harvard-Smithsonian Center for Astrophysics}{HEAD}{Cambridge}{MA}{02138}{USA}
\paperauthor{Kris~Bukovi}{kbukovi@cfa.harvard.edu}{0000-0002-5827-2434}{Harvard-Smithsonian Center for Astrophysics}{HEAD}{Cambridge}{MA}{02138}{USA}
\paperauthor{Nathan~Rapport}{nrapport@cfa.harvard.edu}{0000-0003-4747-8055}{Harvard-Smithsonian Center for Astrophysics}{HEAD}{Cambridge}{MA}{02138}{USA}

  
\begin{abstract}

The new NASA Astrophysics Data System (ADS) is designed with a service-oriented architecture (SOA) that consists of multiple customized Apache Solr search engine instances plus a collection of microservices, containerized using Docker, and deployed in Amazon Web Services (AWS). For complex systems, like the ADS, this loosely coupled architecture can lead to a more scalable, reliable and resilient system if some fundamental questions are addressed. After having experimented with different AWS environments and deployment methods, we decided in December 2017 to go with Kubernetes as our container orchestration. Defining the best strategy to properly setup Kubernetes has shown to be challenging: automatic scaling services and load balancing traffic can lead to errors whose origin is difficult to identify, monitoring and logging the activity that happens across multiple layers for a single request needs to be carefully addressed, and the best workflow for a Continuous Integration and Delivery (CI/CD) system is not self-evident. We present here how we tackle these challenges and our plans for the future.
  
\end{abstract}

\section{Introduction}

The NASA Astrophysics Data System \citep[ADS;][]{2000A&AS..143...41K} is a key bibliographic service for astronomical research. ADS content has steadily increased since its early years \citep{2000A&AS..143..111G}, containing now more than 13 million records and 100 million citations including software and data citations \citep{2015scop.confE...3A}. After several iterations, its original architecture \citep{2000A&AS..143...85A} and user interface \citep{2000A&AS..143...61E} have evolved to address growing maintenance challenges and to adopt newer technologies that allow more advanced functionality \citep{2015ASPC..495..401C, 2015ASPC..492..189A, 2018AAS...23136217A}.


The new ADS is designed with a service-oriented architecture (SOA), containerized using Docker\footnote{\url{https://www.docker.com/}}, orchestrated by Kubernetes\footnote{\url{https://kubernetes.io/}} and deployed in Amazon Web Services\footnote{\url{https://aws.amazon.com/}} (AWS). We have been using this platform for almost a year now, both in our development and production environments. However, when searching for Kubernetes in the full text of the astronomy collection in the new ADS, we currently find only nine results and one of them is not related to the software platform. Among these results, only three present results or a product/service that used Kubernetes in production \citep{2018arXiv180103181A, 2018A&C....25..110A, 2018SPIE10707E..2RF}. The rest only mention the software as an alternative or indicate they are considering to migrate their platform to it in the future. While the new ADS does not have full text for all records, these data indicate that the new ADS is using cutting edge technology in production. The price to pay for being early-adopters is the challenge of solving problems that nobody (or very few people) has faced yet, but sharing our experience will ease the path for others while ADS continues to lead the way in the astrophysical community.

\section{The new architecture}

The new ADS consists of multiple customized Apache Solr\footnote{\url{https://lucene.apache.org/solr/}} search engine instances plus a collection of microservices deployed in two different Kubernetes clusters (see Figure~\ref{fig1}). This loosely coupled architecture allows us to have a more scalable, reliable and resilient system. 

\articlefigure{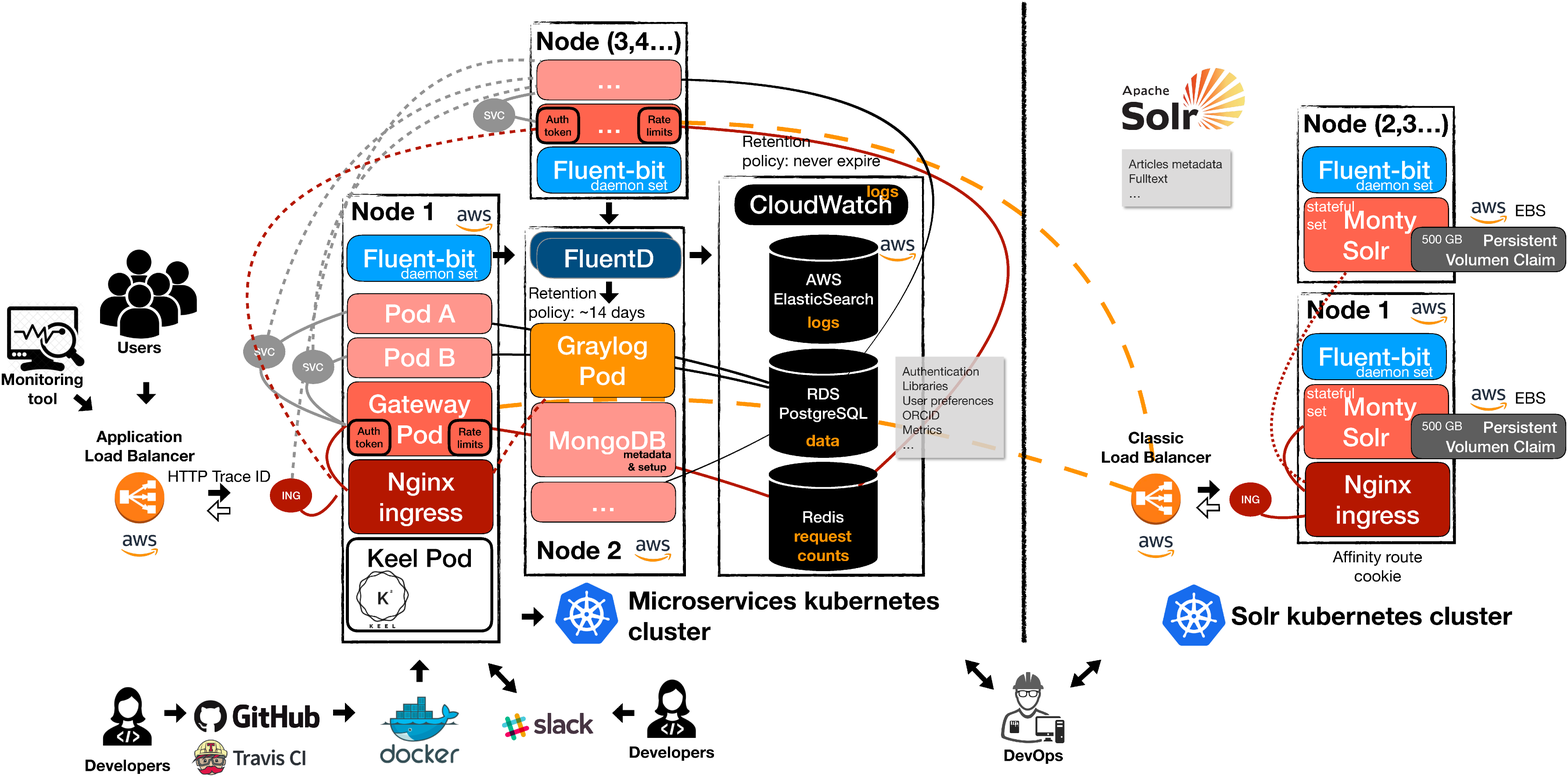}{fig1}{The new ADS architecture with two Kubernetes clusters running the microservices behind the API and an improved Apache Solr search engine.}

Based on our experience, managing Kubernetes clusters in production requires a good strategy to properly monitor all the services from the exterior, log the internal events triggered by users' requests and define a solid strategy to deploy new software versions with a workflow for a Continuous Integration and Delivery (CI/CD) that minimizes service interruptions.

\subsection{Monitoring}

Making sure the whole system is healthy and responding to users' requests is a priority. We developed a custom monitoring tool that emulates users' behavior (e.g., executing searches, accessing libraries, exporting records, filtering results) and alerts us to unexpected results or errors via Slack\footnote{\url{https://slack.com/}}. This emulation happens with a high cadence of the order of several minutes. Historical data is also accumulated and daily reports are generated to measure trends and improvements that could be correlated with microservices updates or infrastructure changes.

\subsection{Logging}

Responding to a single user request may involve multiple microservices (e.g., libraries, Solr search service) and different data requests (e.g., bibcodes in a library, records in Solr). At the very first step, when the user request reaches the AWS application load balancer, a trace identifier is attached to the HTTP request and we propagate it for each required internal request inside our infrastructure. All the microservices output logs to stdout, including key information such as the trace identifier and the user's account identifier. Logs are captured by Fluent Bit\footnote{\url{https://fluentbit.io/}} and distributed to Graylog\footnote{\url{https://www.graylog.org/}} and AWS CloudWatch via Fluentd\footnote{\url{https://www.fluentd.org/}}.  

\subsection{Deploying}

The deployment of new microservice releases is automatically managed by Keel\footnote{\url{https://keel.sh/}}. The developers push new commits to GitHub\footnote{\url{https://github.com/}} and/or make releases, which triggers unit testing via Travis\footnote{\url{https://travis-ci.org/}} continuous integration and image building via Docker hub\footnote{\url{https://hub.docker.com/}}. When a new image is built, Keel deploys it directly to our development environment (each pushed commit) or to our quality assurance environment (each new release). Confirmation to deploy a release in production is provided via Slack, where Keel reports its operations and reacts to developers' approvals.

\section{Future plans}

Several microservices still require manual intervention in order to deploy new releases, Keel does not cover all our development cases and we are working on a new custom tool to meet our needs (after having discarded other tools available in the market due to their complexity). We seek to fully automate the deployment process, while ensuring traceability and easy roll-backs based on automatic functional tests from our monitoring tool. Additionally, to reduce the required resources and simplify operations, we will evaluate other engines for searching through our logs such as Kibana via ElasticSearch\footnote{\url{https://www.elastic.co/products/kibana}} (provided by AWS).

\bibliography{P6-1}

\end{document}